\documentclass[twocolumn,pra,showpacs]{revtex4-1}
\usepackage{amsfonts}
\usepackage{amssymb}
\usepackage{amsmath}
\usepackage{graphicx}
\usepackage{epsfig}

\input{tcilatex}

\begin{document}

\title{Topological Characterization of Extended Quantum Ising Models}
\author{G. Zhang and Z. Song}
\email{songtc@nankai.edu.cn}

\begin{abstract}
We show that a class of exactly solvable quantum Ising models, including the
transverse-field Ising model and anisotropic $XY$ model, can be
characterized as the loops in a two-dimensional auxiliary space. The
transverse-field Ising model corresponds to a circle and the $XY$ model
corresponds to an ellipse, while other models yield cardioid, limacon,
hypocycloid, and Lissajous curves etc. It is shown that the variation of
the ground state energy density, which is a function of the loop,
experiences a nonanalytical point when the winding number of the
corresponding loop changes. The winding number can serve as a topological
quantum number of the quantum phases in the extended quantum Ising model,
which sheds some light upon the relation between quantum phase transition
and the geometrical order parameter characterizing the phase diagram.
\end{abstract}

\maketitle

\affiliation{School of Physics, Nankai University, Tianjin 300071, China}


\textit{Introduction.} Characterizing the quantum phase transitions (QPTs)
is of central significance to both condensed matter physics and quantum
information science. QPTs occur only at zero temperature due to the
competition between different parameters describing the interactions of the
system. A quantitative understanding of the second-order QPT is that the
ground state undergoes qualitative changes when an external parameter passes
through quantum critical points (QCPs).

There are two prototypical models, Bose-Hubbard model and transverse-field
Ising model, based on which the concept and characteristic of\ QPTs can be
well demonstrated. However, among the two, only the transverse-field Ising
model is exactly solvable \cite{SachdevBook}, so as to be a unique paradigm
for understanding the QPTs. Recently, more attention has been paid to
theoretical studies of exactly solvable quantum spin models involving
nearest-, next-nearest-neighbor interactions, and multiple spin exchange
models, etc \cite%
{Titvinidze03,Tsvelik90,Frahm92,Muramoto99,Zvyagin01,Zvyagin03,Lou04,Zvyagin05}%
. Those models are closer to real quasi-one-dimensional magnets \cite%
{Zheludev00,Tsukada00,Kohgi01} comparing to standard ones with only
nearest-neighbor couplings. Furthermore, it has been shown that quantum spin
models can be simulated in artificial quantum system with controllable
parameters. Quantum simulation of spin chain can be experimentally realized
through neutral atoms stored in an optical lattice \cite{Simon11,Struck11},
trapped ions \cite%
{Porras04,Deng05,Taylor08,Friedenauer08,Kim10,Edwards10,Kim11,Kim09,Islam11}
and NMR simulator \cite{Li14}.\ This system often serves as a test bed for
applying new ideas and methods to quantum phase transitions.

A fundamental question is whether QPTs in Ising model can have a connection
to some topological characterizations. It is interesting to note in this
context that some simple Ising models have been found to exhibit topological
characterization \cite{Lee07,Feng07,DeGottardi11,Zhang11}.\ The purpose of
the present work is to shed some light upon the relation between QPTs and a
geometrical parameter characterizing the phase diagram, through the
investigation of a class of quantum Ising models.

In this work, we present an extended quantum Ising model, which includes an
additional three-body interaction. It can be exactly solved by the routine
procedure, taking the Jordan-Wigner and pseudo-spin transformations. Based
on the exact solution, we investigate the QPT in this model. We introduce a
global order parameter, which is the winding number for the loop specifying
to a set of coupling constants, in an auxiliary space. The ground state
energy density can be a function of the loop and its variation experiences a
nonanalytical point when the winding number of the corresponding loop
changes. Then the relation between QPTs and the geometrical order parameter
is established.

\textit{Extended Ising model and solutions. }We start our analysis from the
one-dimensional Ising model, which has the Hamiltonian

\begin{eqnarray}
H &=&\sum\limits_{j=1}^{N}[a\left( \frac{1+\gamma }{2}\sigma _{j}^{x}\sigma
_{j+1}^{x}+\frac{1-\gamma }{2}\sigma _{j}^{y}\sigma _{j+1}^{y}\right)
+g\sigma _{j}^{z}  \notag \\
&&+b\sigma _{j}^{z}\left( \frac{1+\delta }{2}\sigma _{j-1}^{x}\sigma
_{j+1}^{x}+\frac{1-\delta }{2}\sigma _{j-1}^{y}\sigma _{j+1}^{y}\right) ],
\label{H}
\end{eqnarray}%
where $\sigma _{j}^{\alpha }$, for $\alpha =x,y,z,$ are the usual Pauli
matrices, and periodic boundary conditions are assumed. Comparing with the
customary anisotropic $XY$\ model, there are additional three-site
interactions $\sigma _{j}^{z}\sigma _{j-1}^{x}\sigma _{j+1}^{x}$ and $\sigma
_{j}^{z}\sigma _{j-1}^{y}\sigma _{j+1}^{y}$, which have\ the following\ two
implications: it can be either regarded as the conditional anisotropic $XY$%
-type coupling between next-nearest-neighbor spins, or conditional action of
transverse field. The ground state phase diagram and correlation functions
for this spin model have been studied 40 years ago \cite{Suzuki71}.\ In the
case of $g=\gamma =\delta =0$, the correlation function has been obtained
\cite{Titvinidze03}. In addition, other types of Hamiltonians which contain
three-body interactions were also investigated \cite%
{Zvyagin06,Divakaran13,Zhang12}.

\begin{figure}[tbp]
\begin{center}
\includegraphics[bb=40 360 440 780, width=0.20\textwidth, clip]{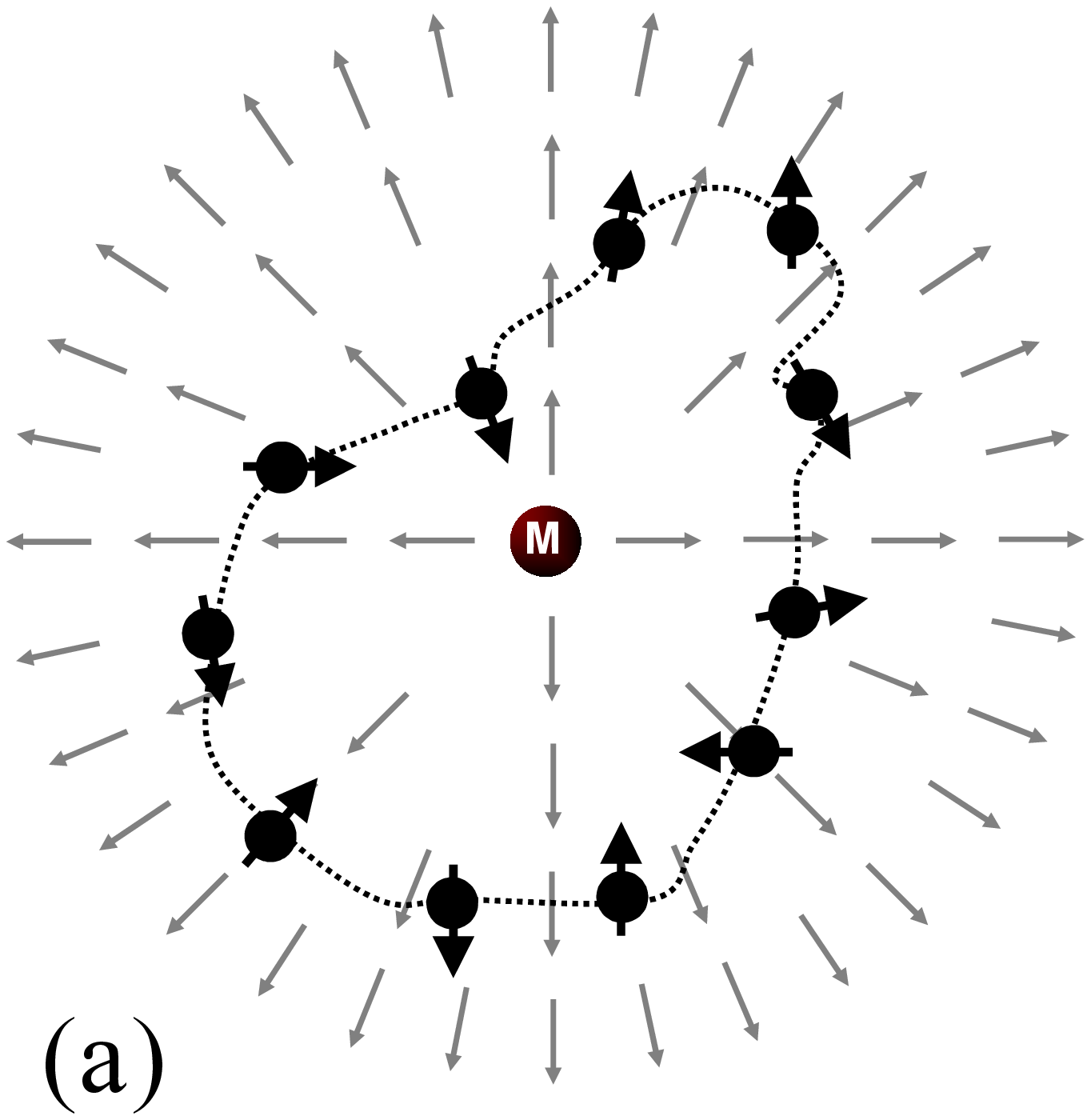}
\includegraphics[bb=50 388 420 770, width=0.20\textwidth, clip]{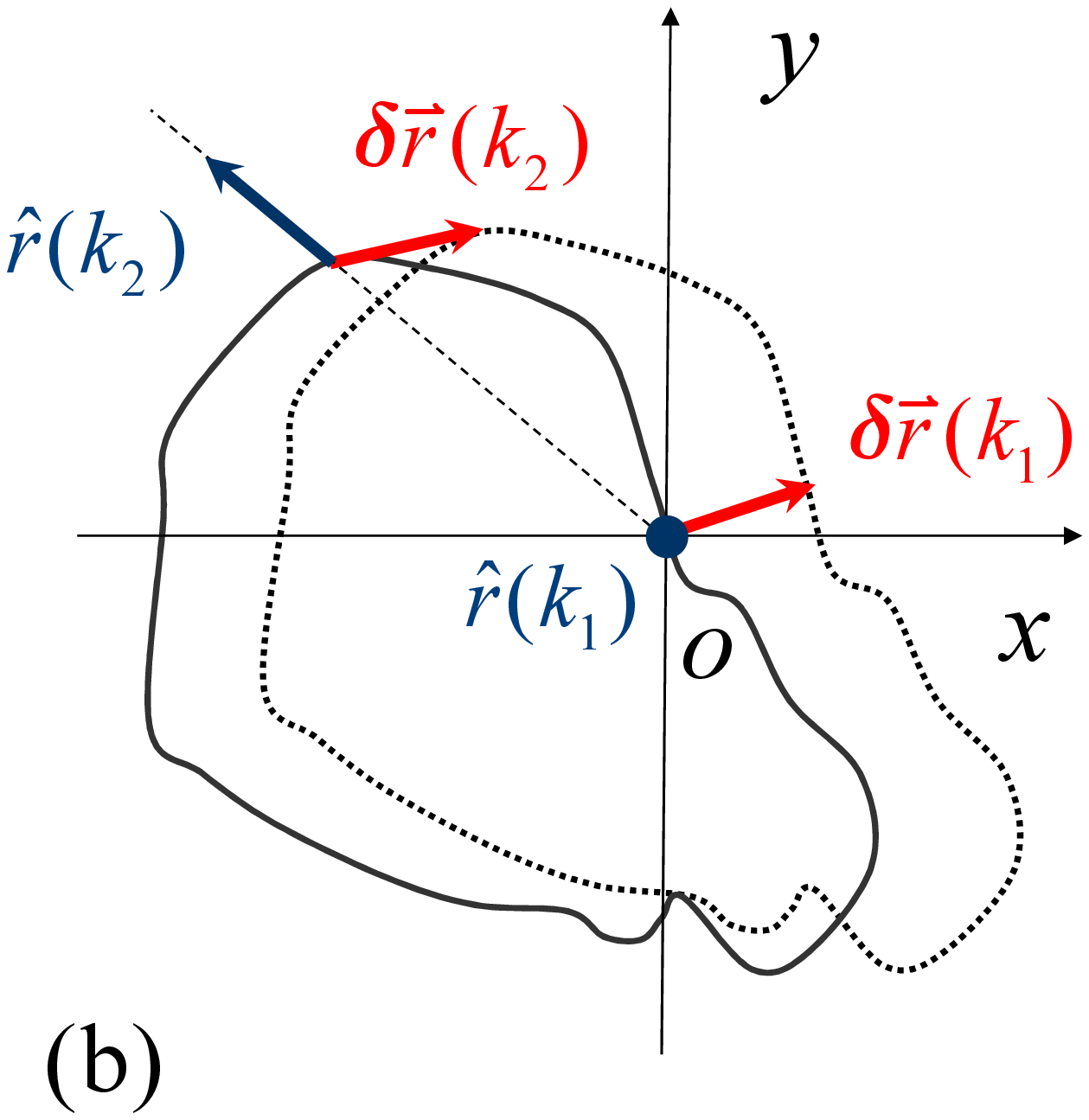}
\end{center}
\caption{(Color online) Schematically illustration of the equivalent
Hamiltonian (\protect\ref{H_eq}), which represents a system of spin ensemble
in a monopole field. In the thermodynamic limit, the ground state energy
density becomes an integration corresponding to a loop. (b) Schematics of
two loops $l$ and $l^{\prime }$ described by the parametric equations
involving vectors $\protect\overrightarrow{r}\left( k\right) $\ and $\protect%
\overrightarrow{r}\left( k\right) +\protect\delta \protect\overrightarrow{r}%
\left( k\right) $, respectively. Here $\protect\overrightarrow{r}\left(
k_{1}\right) $ represents an arbitrary point on the $l$, while $\protect%
\overrightarrow{r}\left( k_{1}\right) +\protect\delta \protect%
\overrightarrow{r}\left( k_{1}\right) $ represents the corresponding point
on the $l^{\prime }$. The red arrow indicates $\protect\delta \protect%
\overrightarrow{r}\left( k_{1}\right) $, while the blue arrow\ indicates the
corresponding unitary vector $\hat{r}\left( k_{1}\right) $. The inner
product between them $\protect\delta \protect\overrightarrow{r}\left(
k_{1}\right) \cdot \hat{r}\left( k_{1}\right) $\ contributes to the
variation of $\protect\varepsilon _{g}$. The loop $l$\ passes the origin at
point $\protect\overrightarrow{r}\left( k\right) $ with $k=k_{0}$. The
corresponding unitary vector $\hat{r}\left( k_{0}\right) $\ is indefinite,
which is denoted as a solid blue circle. The indefiniteness of $\protect%
\delta \protect\overrightarrow{r}\left( k_{0}\right) \cdot \hat{r}\left(
k_{0}\right) $\ witnesses the QPT as well as the topological change of the
loop: enclosing the origin or not.}
\label{figure1}
\end{figure}

\begin{figure}[tbp]
\begin{center}
\includegraphics[bb=170 270 420 540, width=0.13\textwidth, clip]{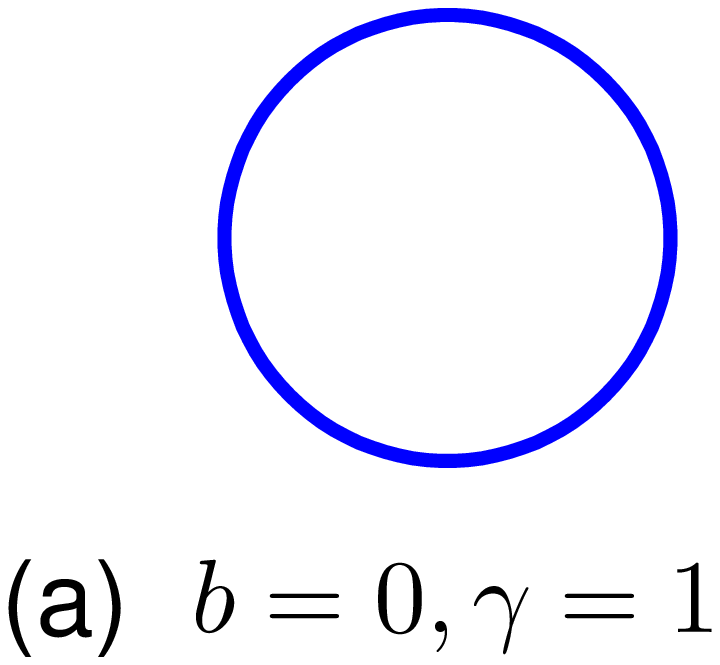} %
\includegraphics[bb=170 270 420 540, width=0.13\textwidth, clip]{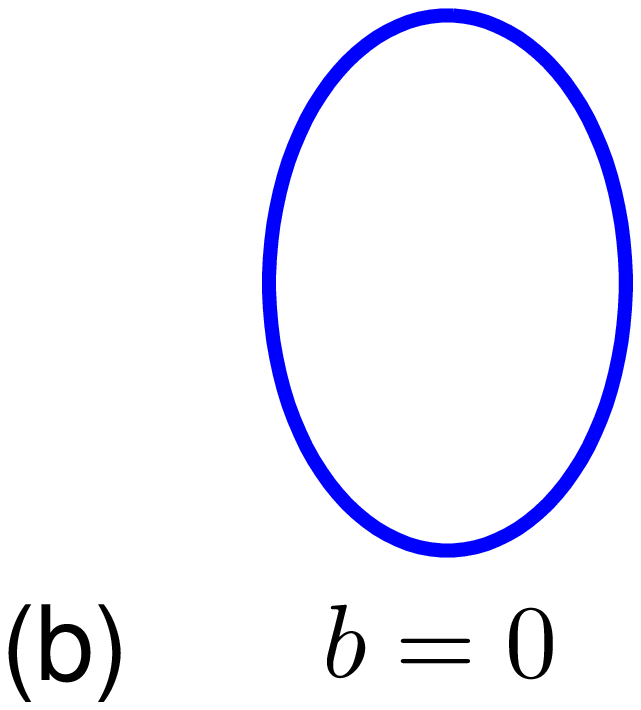}
\includegraphics[bb=170 270 420 540, width=0.13\textwidth,
clip]{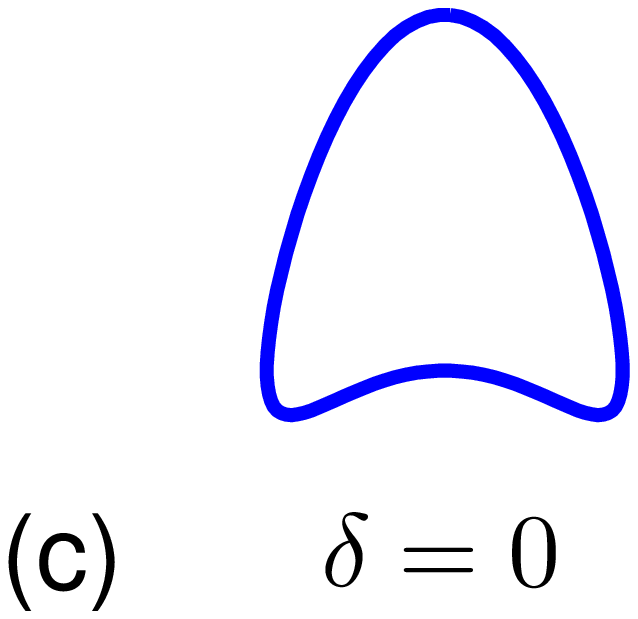}
\includegraphics[bb=170 270 420 540,
width=0.13\textwidth, clip]{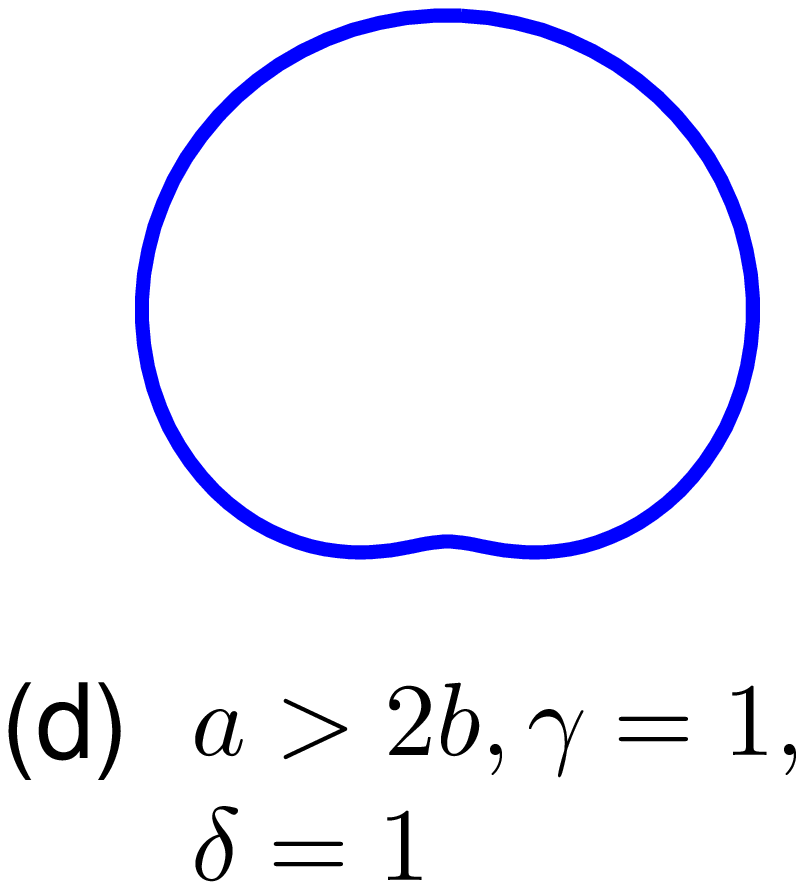}
\includegraphics[bb=170 270 420
540, width=0.13\textwidth, clip]{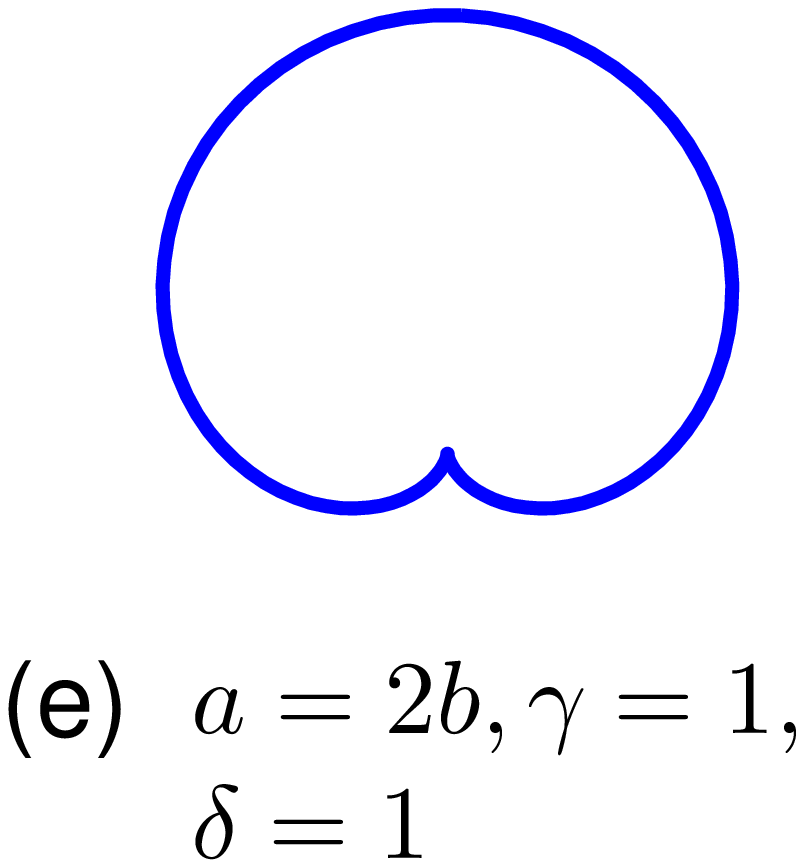}
\includegraphics[bb=170 270
420 540, width=0.13\textwidth, clip]{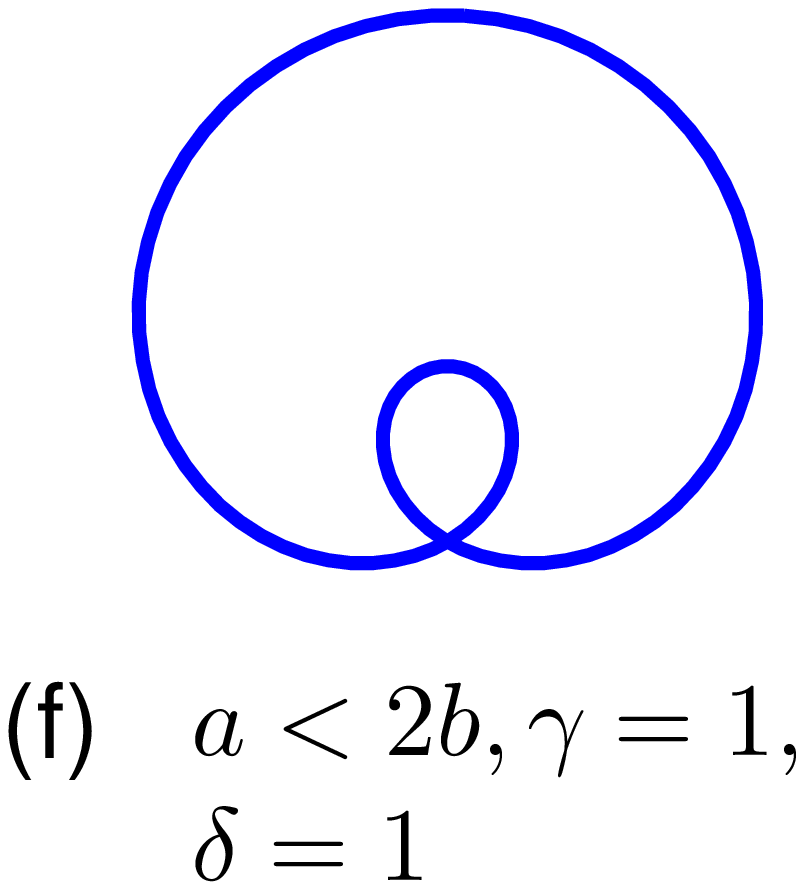}
\includegraphics[bb=170
270 420 540, width=0.13\textwidth, clip]{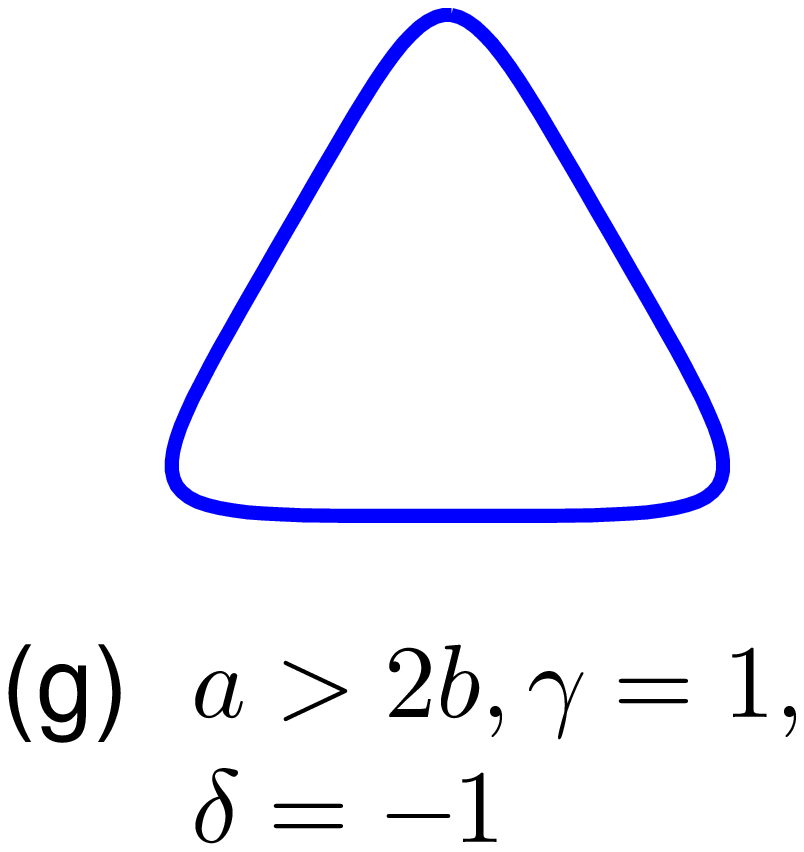}
\includegraphics[bb=170 270 420 540, width=0.13\textwidth,
clip]{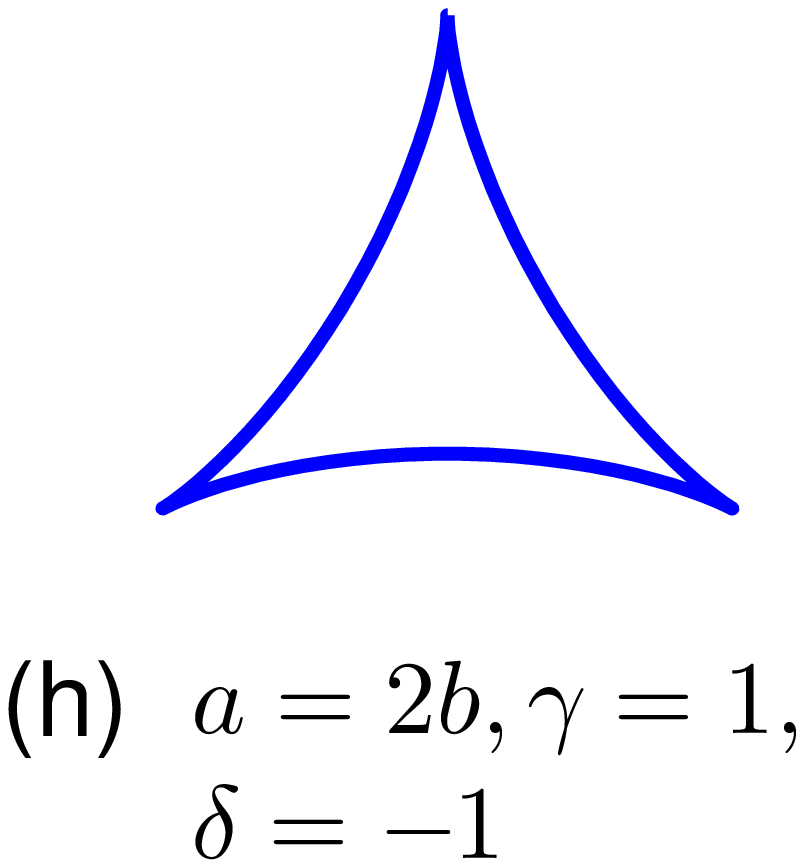}
\includegraphics[bb=170 270 420 540,
width=0.13\textwidth, clip]{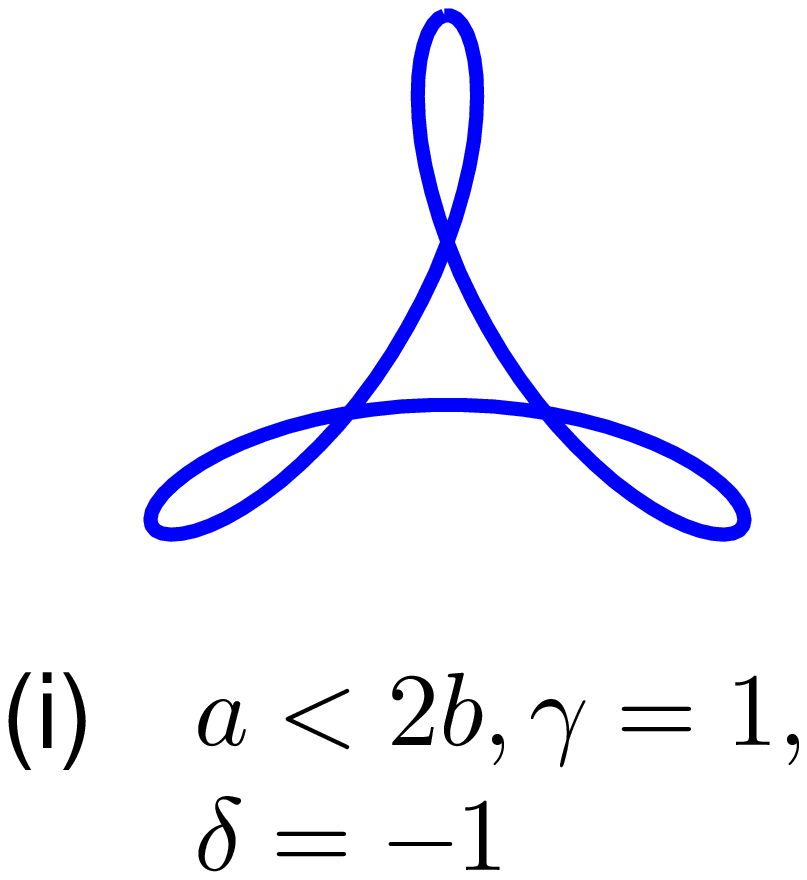}
\includegraphics[bb=170 270
420 540, width=0.13\textwidth, clip]{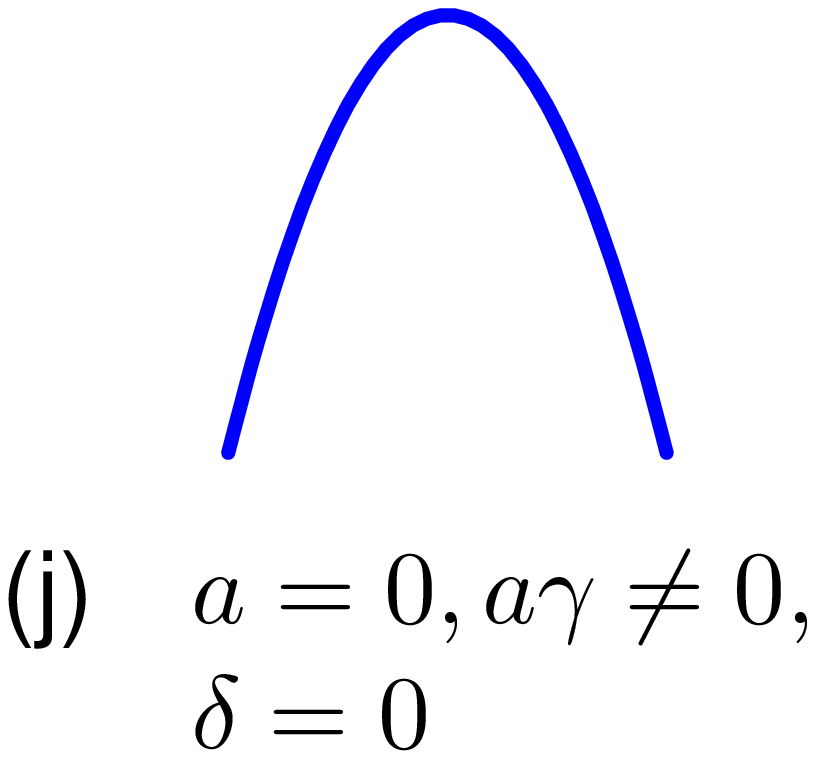}
\includegraphics[bb=170
270 420 540, width=0.13\textwidth, clip]{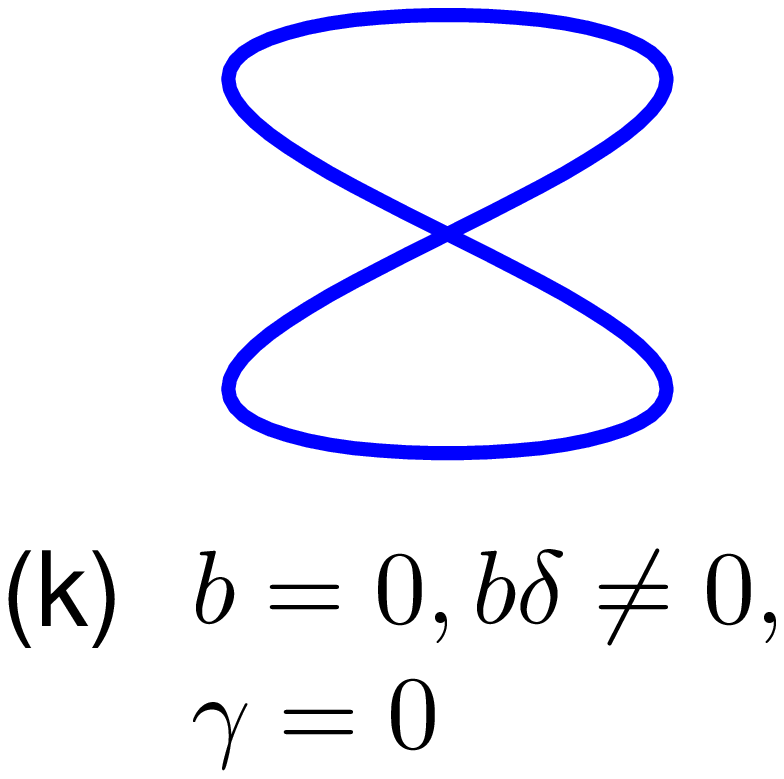} %
\includegraphics[bb=170 270 420 540, width=0.13\textwidth, clip]{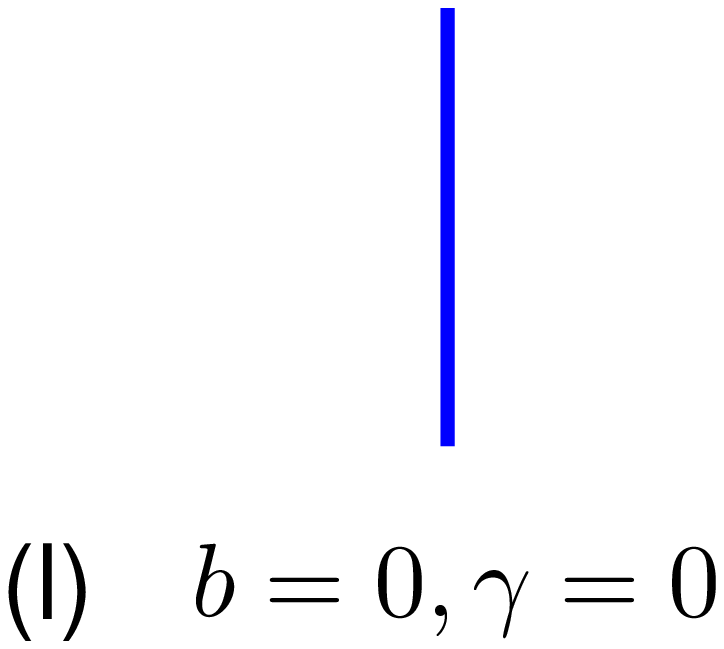}
\end{center}
\caption{(Color online) several types of graphs}
\label{figure2}
\end{figure}

We will see that this model can be exactly solvable in a simple way by the
similar procedure for the simple transverse-field Ising model \cite%
{SachdevBook,Dziarmaga,Pfeuty}. For the sake of simplicity, we only concern
the case of even $N$, the conclusion is available for the case of odd $N$ in
the thermodynamic limit. As the same procedure performed in solving the
Hamiltonian without the additional term, we take the Jordan-Wigner
transformation \cite{SachdevBook}%
\begin{eqnarray}
\sigma _{j}^{z} &=&1-2c_{j}^{\dagger }c_{j}\text{, }\sigma _{j}^{y}=\mathrm{i%
}\sigma _{j}^{x}\sigma _{j}^{z}, \\
\sigma _{j}^{x} &=&-\prod\limits_{l<j}\left( 1-2c_{l}^{\dagger }c_{l}\right)
\left( c_{j}+c_{j}^{\dag }\right) ,  \label{JW}
\end{eqnarray}%
to replace the Pauli operators by the fermionic operators $c_{j}$. We note
that the parity of the number of fermions is a conservative quantity and\
then the Hamiltonian (\ref{H}) can be written in the form
\begin{equation}
H=\left(
\begin{array}{cc}
H^{+} & 0 \\
0 & H^{-}%
\end{array}%
\right) ,  \label{H_m}
\end{equation}%
where%
\begin{eqnarray}
H^{+} &=&H^{-}-2\left[ b\left( c_{N}^{\dagger }c_{2}+c_{1}^{\dagger
}c_{N-1}+\delta c_{2}c_{N}+\delta c_{1}c_{N-1}\right) \right.  \notag \\
&&\left. +a\left( c_{N}^{\dag }c_{1}+\gamma c_{1}c_{N}\right) +\mathrm{H.c.}%
\right]  \label{H+}
\end{eqnarray}%
and%
\begin{eqnarray}
H^{-} &=&\sum\limits_{j=1}^{N}\left[ \left( g/2-gc_{j}^{\dagger
}c_{j}\right) +a\left( c_{j}^{\dag }c_{j+1}+\gamma c_{j+1}c_{j}\right)
\right.  \notag \\
&&\left. +b\left( c_{j}^{\dagger }c_{j+2}+\delta c_{j+2}c_{j}\right) \right]
+\mathrm{H.c.}  \label{H-}
\end{eqnarray}%
are corresponding reduced Hamiltonians in the invariant subspaces with even
and odd number of fermions. Here $H^{+}$\ represents a fermionic ring
threaded by a half of the flux quantum.\ In the following, we will focus on $%
H^{+}$\ since the ground state has even parity for any values of parameters.
Similarly, $H^{+}$ can be diagonalized by Fourier and pseudo-spin
transformations. Taking the Fourier transformation%
\begin{equation}
c_{j}=\frac{1}{\sqrt{N}}\sum_{k}c_{k}e^{ikj},  \label{Fourier}
\end{equation}%
where $k=2\pi \left( m+1/2\right) /N,m=0,1,2,\ldots ,N-1,$ the Hamiltonian $%
H^{+}$\ can be expressed as a compact form

\begin{eqnarray}
H^{+} &=&4\sum\limits_{k>0}\overrightarrow{r}\left( k\right) \cdot
\overrightarrow{s}_{k},  \label{H_eq} \\
\overrightarrow{r}\left( k\right) &=&\left( 0,a\gamma \sin k+b\delta \sin
2k,a\cos k+b\cos 2k-g\right) ,\text{\ \ \ \ \ }  \label{loop}
\end{eqnarray}%
which represents a set of pseudo spins $\left\{ \overrightarrow{s}%
_{k}\right\} $\ in a two-dimensional magnetic field $\overrightarrow{r}$.
The pseudo spin is defined as%
\begin{eqnarray}
s_{k}^{-} &=&\left( s_{k}^{+}\right) ^{\dag }=c_{k}c_{-k}, \\
s_{k}^{z} &=&\frac{1}{2}\left( c_{k}^{\dag }c_{k}+c_{-k}^{\dag
}c_{-k}-1\right) ,  \label{pseudospin}
\end{eqnarray}%
satisfying the SU(2) algebra $\left[ s_{k}^{z},s_{k^{\prime }}^{\pm }\right]
=\pm \delta _{kk^{\prime }}s_{k^{\prime }}^{\pm }$, $\left[
s_{k}^{+},s_{k^{\prime }}^{-}\right] =2\delta _{kk^{\prime }}s_{k^{\prime
}}^{z}.$ It is clear that the equivalent Hamiltonian (\ref{H_eq}) represents
a system of spin ensemble in a monopole field. These spins locate at the
points on the loop of $\overrightarrow{r}\left( k\right) $\ defined in Eq. (%
\ref{loop}). In the following argument, we do not restrict the shape of the
loop. The obtained result is valid for an arbitrary loop, which is
schematically illustrated in \ref{figure1}(a). The Hamiltonian (\ref{H_eq})\
is easy to be diagonalized by aligning all spins with the local magnetic
field, which is the essential of the Bogoliubov transformation. In the
thermodynamic limit, the ground state energy density can be expressed by an
integration

\begin{equation}
\varepsilon _{g}=\lim_{N\rightarrow \infty }\frac{E_{g}}{N}=-\frac{1}{2\pi }%
\int\limits_{-\pi }^{\pi }\left\vert \overrightarrow{r}\left( k\right)
\right\vert \mathrm{d}k,  \label{E/N}
\end{equation}%
which corresponds to a loop tracing with the parametric equation $%
\overrightarrow{r}\left( k\right) =\left( 0,x\left( k\right) ,y\left(
k\right) \right) $. In our case, the parametric equation has the form
\begin{equation}
\left\{
\begin{array}{c}
x\left( k\right) =a\gamma \sin k+b\delta \sin \left( 2k\right) \\
y\left( k\right) =a\cos k+b\cos \left( 2k\right) -g%
\end{array}%
\right.  \label{PE}
\end{equation}%
in the auxiliary space $\left( x,y\right) $. Then we can use some simple
loops to represent the ground state of the Ising-like model. It offers many
types of graphs corresponding different kinds of Ising model. To demonstrate
this point, we plot several types of graphs in Fig. \ref{figure2}. It shows
that two familiar models, transverse Ising model and anisotropic $XY$ model,
correspond to two simple graphs, circle and ellipse, respectively. Rest
models connect to more complicated graphs. Furthermore, the ground state of
each model is naturally connected to a graph individually.

\textit{Quantum phase transition. }In this section, we investigate the QPT
occurring in the extended Ising model and its connection to the geometry of
the corresponding loops. To this end, we start with the change of the ground
state energy induced by varying the parameters. In general the QPT driven by
the parameters $\left\{ \alpha \right\} $ $\left( \alpha =a,b,\gamma
,\lambda ,g\right) $ can be characterized by the derivative of the ground
state density with respect to $\alpha $, $\partial \varepsilon _{g}/\partial
\alpha $ which experiences a nonanalytic point at the critical point,
leading to the divergence of second derivative of ground state energy
density. We investigate the signature of the QPT in an alternative way: $%
\varepsilon _{g}$ depends on the path of the integral, being a function of
the functions $x\left( k\right) $ and $y\left( k\right) $. The parameters $%
\left\{ \alpha \right\} $\ drive the QPT through the change of the functions
$x\left( k\right) $ and $y\left( k\right) $, or the loop. In other words,
one can consider the variations of functions $x\left( k\right) $ and $%
y\left( k\right) $ instead of the change of the parameters $\left\{ \alpha
\right\} $.\ The first variation of the function $\varepsilon _{g}[x,y]$ is%
\begin{equation}
\delta \varepsilon _{g}=\int \left( \frac{\delta \varepsilon _{g}}{\delta x}%
\delta x+\frac{\delta \varepsilon _{g}}{\delta y}\delta y\right) \mathrm{d}%
k=-\frac{1}{2\pi }\int\limits_{-\pi }^{\pi }\hat{r}\left( k\right) \cdot
\delta \overrightarrow{r}\left( k\right) \mathrm{d}k,
\end{equation}%
where $\hat{r}\left( k\right) =\overrightarrow{r}/\left\vert \overrightarrow{%
r}\right\vert $ is the unitary vector of $\overrightarrow{r}\left( k\right) $%
. It indicates that the variation $\delta \varepsilon _{g}$\ is the
summation of the path shifts $\delta \overrightarrow{r}\left( k\right) $\
along the direction of $\hat{r}\left( k\right) $. We are interested in the
case of the loop crossing the origin. At the origin, the unitary vector $%
\hat{r}\left( k\right) $\ is indefinite, which leads to an indefinite
contribution to the variation $\delta \varepsilon _{g}$, indicating a
nonsmooth point. It is a signature of the QPT associated with a topological
change in the loop of the integration. So far, we do not specify the shape
of the loop and how the loop is deformed. In our case, the variation $\delta
\overrightarrow{r}\left( k\right) $ arises from the continuous change of the
parameters $\left\{ \alpha \right\} $. Then we have%
\begin{equation}
\delta \overrightarrow{r}\left( k\right) =\sum_{\alpha }\frac{\partial
\overrightarrow{r}\left( k\right) }{\partial \alpha }\mathrm{d}\alpha
\end{equation}%
or explicitly%
\begin{equation}
\left\{
\begin{array}{c}
\delta x\left( k\right) =\sin k\left( \gamma \mathrm{d}a+a\mathrm{d}\gamma
\right) +\sin \left( 2k\right) \left( b\mathrm{d}\lambda +\lambda \mathrm{d}%
b\right) \\
\delta y\left( k\right) =\cos k\mathrm{d}a+\cos \left( 2k\right) \mathrm{d}b-%
\mathrm{d}g%
\end{array}%
\right. .
\end{equation}

Considering the case with $a=\gamma =1$, $b=\lambda =0$ as an example, the
Hamiltonian (\ref{H}) reduces to the simplest transverse field Ising model%
\begin{equation}
H_{\text{\textrm{Ising}}}=\sum\limits_{j=1}^{N}\left( \sigma _{j}^{x}\sigma
_{j+1}^{x}+g\sigma _{j}^{z}\right) ,
\end{equation}%
the ground state energy density of which corresponds to a circle of the
equation%
\begin{equation}
x^{2}+\left( y+g\right) ^{2}=1.
\end{equation}%
The variation $\delta \varepsilon _{g}$ from the case with $g=\pm 1$ is
readily expressed as%
\begin{equation}
\delta \varepsilon _{g}=-\frac{\mathrm{d}g}{2\pi }\int\limits_{-\pi }^{\pi }%
\left[ \hat{r}\left( k\right) \cdot \hat{\jmath}\right] \mathrm{d}k,
\end{equation}%
where $\hat{\jmath}$\ denotes the unit vector of $y$ axis. It results in $%
\delta \varepsilon _{g}=\partial \varepsilon _{g}/\partial g\mathrm{d}g$,
which shows that $\delta \varepsilon _{g}$ accords with $\partial
\varepsilon _{g}/\partial g$\ as a witness of QPT.

We would like to point out that each loop contains two characters, geometry
(shape and position) and curve orientation, which are determined by the
corresponding parameter equation. To characterize these two features, we use
a topological quantity, winding number, which is a fundamental concept in
geometric topology and widely used in various areas of physics \cite%
{Park2000,Hatsugai93,Leutwyler92,Ezawa13}. The winding number of a closed
curve in the auxiliary $xy$-plane around the origin is defined as

\begin{equation}
\mathcal{N}=\frac{1}{2\pi }\int\nolimits_{c}\frac{1}{r^{2}}\left( y\mathrm{d}%
x-x\mathrm{d}y\right) ,  \label{winding N}
\end{equation}%
which is an integer representing the total number of times that the curve
travels clockwise around the origin. Then we establish the connection
between the QPT and the switch of the topological quantity.

\begin{figure}[tbp]
\begin{center}
\includegraphics[bb=50 300 520 700, width=0.40\textwidth, clip]{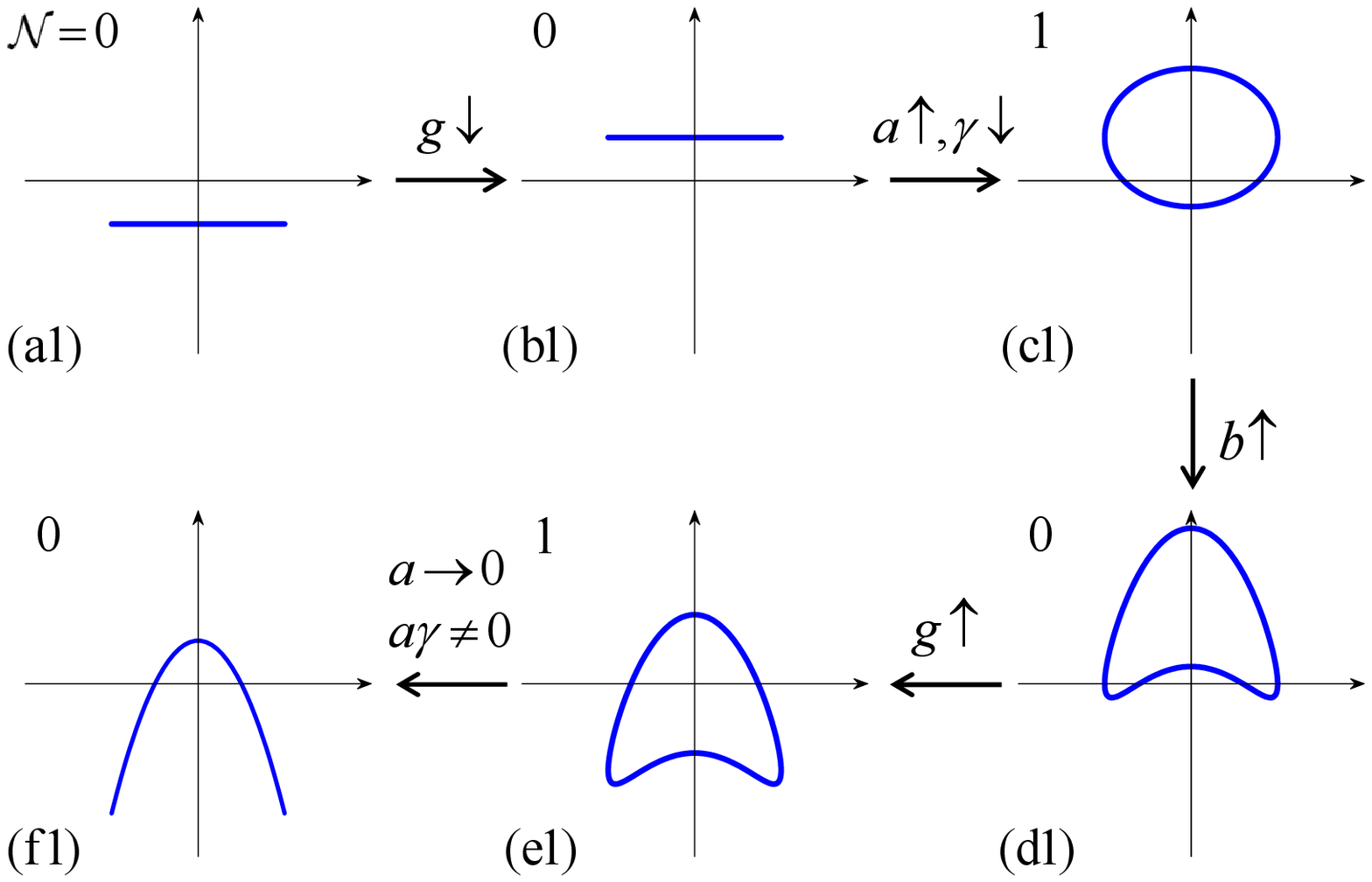} %
\includegraphics[bb=50 300 520 700, width=0.40\textwidth, clip]{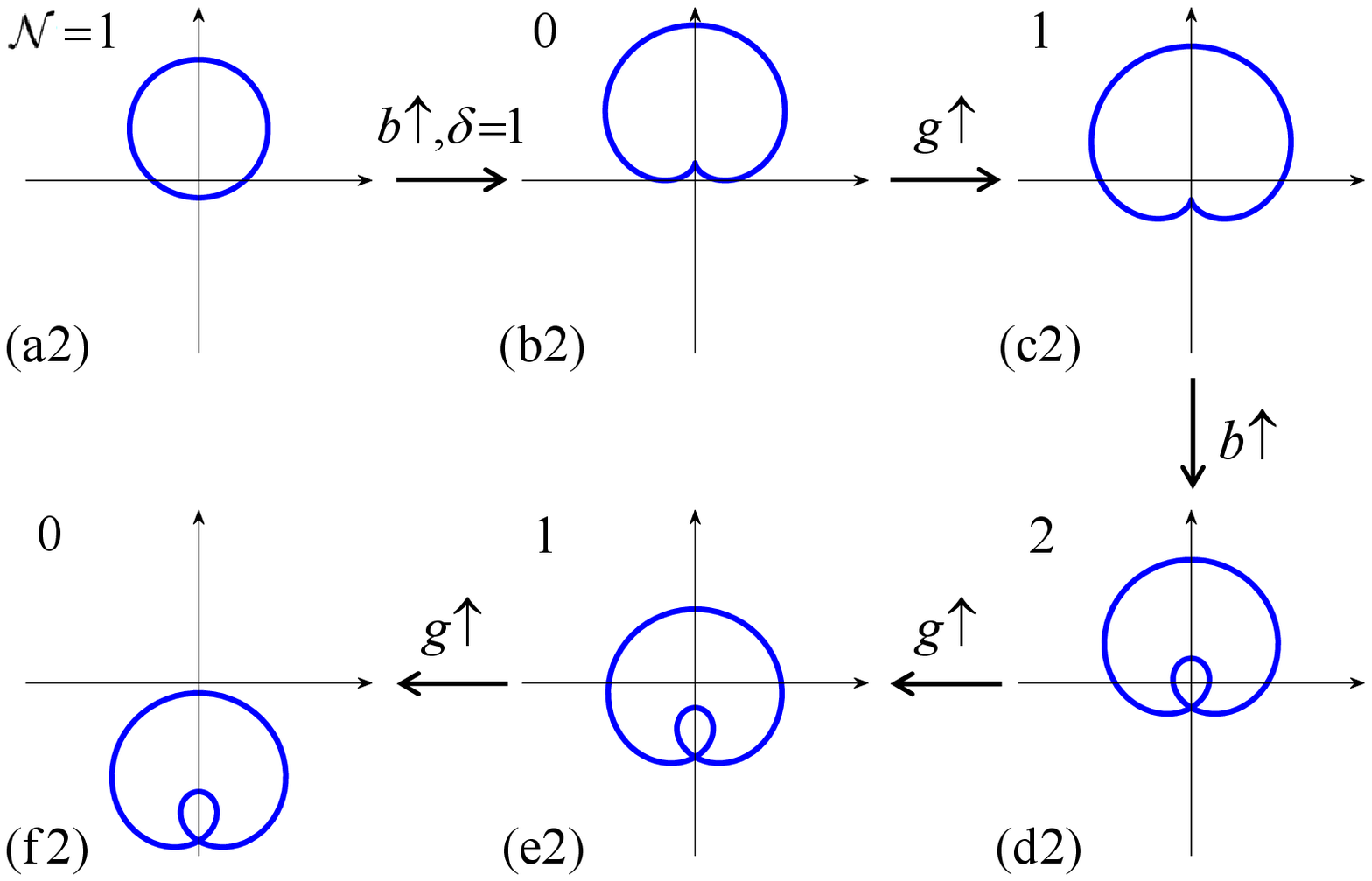} %
\includegraphics[bb=50 360 520 700, width=0.40\textwidth, clip]{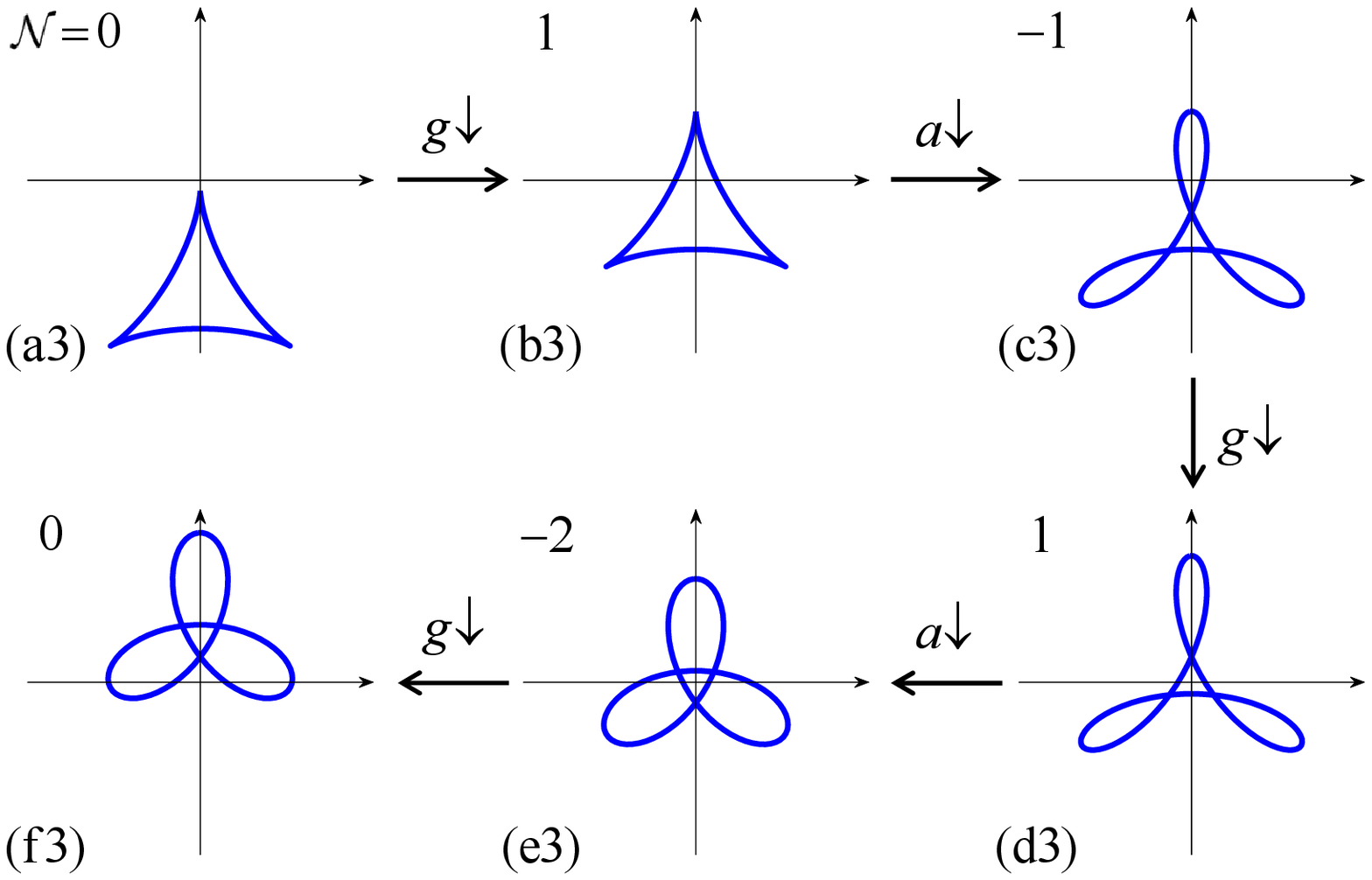}
\end{center}
\caption{(Color online) The winding numbers for various typical cases
corresponding graphs in an auxiliary space with different topologies. Label $%
\uparrow $ ($\downarrow $)\ denotes the increase (decrease) of the
parameters, which induce the transition between graphs.}
\label{figure3}
\end{figure}

\textit{Topological quantum number. }We calculate the winding numbers for
various typical cases corresponding to graphs in an auxiliary space with
different topologies. In Fig. \ref{figure3}, we plot the graphs of the
ground state for typical cases. The corresponding winding number and the
relations between graphs are presented. In each group, the first graph is
clockwise.\ The examples show that there are five possible winding numbers $%
\pm 2$, $\pm 1$, and $0$, which represent five different phases.

To demonstrate the characteristics of these phases, we consider five typical
cases, which correspond to the ground states of systems $h_{\mathcal{N}}$
with parameters in following limits: (i) $b\rightarrow \infty $ and $\delta
=-1$, the reduced Hamiltonian is $h_{-2}=N^{-1}\sum\nolimits_{j=1}^{N}\sigma
_{j}^{z}\sigma _{j-1}^{y}\sigma _{j+1}^{y}$, (ii) $a\rightarrow \infty $ and
$\gamma =-1$ and $h_{-1}=N^{-1}\sum\nolimits_{j=1}^{N}\sigma _{j}^{y}\sigma
_{j+1}^{y}$, (iii) $g\rightarrow \infty $, $h_{0}=N^{-1}\sum%
\nolimits_{j=1}^{N}\sigma _{j}^{z}$, (iv) $a\rightarrow \infty $ and $\gamma
=1$, $h_{1}=N^{-1}\sum\nolimits_{j=1}^{N}\sigma _{j}^{x}\sigma _{j+1}^{x}$,
(v) $b\rightarrow \infty $ and $\delta =1$, $h_{2}=N^{-1}\sum%
\nolimits_{j=1}^{N}\sigma _{j}^{z}\sigma _{j-1}^{x}\sigma _{j+1}^{x}$. The
corresponding ground states $\left\vert G_{0,\pm 1}\right\rangle $ of
even-number flip subspace, represented in position space, are readily
obtained as
\begin{eqnarray}
\left\vert G_{0}\right\rangle &=&\prod_{j}\left\vert \downarrow
\right\rangle _{j}, \\
\left\vert G_{1}\right\rangle &=&\tfrac{1}{\sqrt{2}}(\prod_{j\in \mathrm{e}%
}\left\vert \nearrow \right\rangle _{j}\prod_{j\in \mathrm{o}}\left\vert
\swarrow \right\rangle _{j}+\prod_{j\in \mathrm{e}}\left\vert \swarrow
\right\rangle _{j}\prod_{j\in \mathrm{o}}\left\vert \nearrow \right\rangle
_{j}),\text{ \ \ \ } \\
\left\vert G_{-1}\right\rangle &=&\tfrac{1}{\sqrt{2}}(\prod_{j\in \mathrm{e}%
}\left\vert \searrow \right\rangle _{j}\prod_{j\in \mathrm{o}}\left\vert
\nwarrow \right\rangle _{j}+\prod_{j\in \mathrm{e}}\left\vert \nwarrow
\right\rangle _{j}\prod_{j\in \mathrm{o}}\left\vert \searrow \right\rangle
_{j}),
\end{eqnarray}%
where $\sigma _{j}^{z}\left\vert \downarrow \right\rangle _{j}(\left\vert
\uparrow \right\rangle _{j})=-\left\vert \downarrow \right\rangle
_{j}(\left\vert \uparrow \right\rangle _{j})$, $\sigma _{j}^{x}\left\vert
\nearrow \right\rangle _{j}(\left\vert \swarrow \right\rangle
_{j})=\left\vert \nearrow \right\rangle _{j}(-\left\vert \swarrow
\right\rangle _{j})$, $\sigma _{j}^{y}\left\vert \searrow \right\rangle
_{j}(\left\vert \nwarrow \right\rangle _{j})=\left\vert \searrow
\right\rangle _{j}(-\left\vert \nwarrow \right\rangle _{j})$, and \textrm{e}
and \textrm{o} denote the even and odd number of sites, respectively. The
ground states of $h_{\pm 2}$\ are obtained from Eq. (\ref{H_eq}) and
expressed in an auxiliary space as%
\begin{equation}
\left\vert G_{\pm 2}^{\prime }\right\rangle =\prod_{k>0}\left( \pm \mathrm{i}%
\sin k\left\vert \uparrow \right\rangle _{k}+\cos k\left\vert \downarrow
\right\rangle _{k}\right) ,  \label{G_2_k}
\end{equation}%
where $\left\vert \uparrow \right\rangle _{k}$ and $\left\vert \downarrow
\right\rangle _{k}$ are eigenstates of pseudo-spin operator $s_{k}^{z}$ with
$2s_{k}^{z}\left\vert \uparrow \right\rangle _{k}(\left\vert \downarrow
\right\rangle _{k})=\left\vert \uparrow \right\rangle _{k}(-\left\vert
\downarrow \right\rangle _{k})$. It is a little complicated to express
states $\left\vert G_{\pm 2}\right\rangle $\ in the position space in a
simple form. Here we only give the expression for $N=4n$ ($n\in
\mathbb{N}
$) \cite{Proof},
\begin{eqnarray}
\left\vert G_{\pm 2}\right\rangle &=&2^{-\left( N-2\right)
/2}\sum_{j=0}^{N/2}\left( \pm 1\right) ^{j}e^{i\frac{\pi }{2}%
\sum_{l=1}^{2j}\left( -1\right) ^{l}n_{l}}  \notag \\
&&\times \prod\nolimits_{\left\{ \sum\nolimits_{l=1}^{2j}n_{l}=\mathrm{even}%
\right\} }\sigma _{n_{l}}^{-}\left\vert \Uparrow \right\rangle ,
\label{G_2_R}
\end{eqnarray}%
where $\left\vert \Uparrow \right\rangle =\prod_{l=1}^{N}\left\vert \uparrow
\right\rangle _{l}$ is the saturate ferromagnetic state. As an example, the
ground states for $4$-site systems $h_{\pm 2}$ are explicitly
\begin{eqnarray}
\left\vert G_{\pm 2}^{N=4}\right\rangle &=&1/2(\left\vert \uparrow
\right\rangle _{1}\left\vert \uparrow \right\rangle _{2}\left\vert \uparrow
\right\rangle _{3}\left\vert \uparrow \right\rangle _{4}\mp \left\vert
\uparrow \right\rangle _{1}\left\vert \downarrow \right\rangle
_{2}\left\vert \uparrow \right\rangle _{3}\left\vert \downarrow
\right\rangle _{4}  \notag \\
&&\mp \left\vert \downarrow \right\rangle _{1}\left\vert \uparrow
\right\rangle _{2}\left\vert \downarrow \right\rangle _{3}\left\vert
\uparrow \right\rangle _{4}-\left\vert \downarrow \right\rangle
_{1}\left\vert \downarrow \right\rangle _{2}\left\vert \downarrow
\right\rangle _{3}\left\vert \downarrow \right\rangle _{4}).
\end{eqnarray}

We employ the expected value of operators $h_{\rho }$, $\left\langle
G_{\lambda }\right\vert h_{\rho }\left\vert G_{\lambda }\right\rangle $, as
local order parameters to characterize the ground states $\left\vert
G_{\lambda }\right\rangle $ ($\rho ,\lambda =\pm 2,\pm 1,0$). By using the
similar analysis in \cite{Proof}, we have
\begin{equation}
\left\langle G_{\lambda }\right\vert h_{\rho }\left\vert G_{\lambda
}\right\rangle =-\delta _{\lambda \rho }.
\end{equation}%
It indicates that the five ground states $\left\vert G_{\lambda
}\right\rangle $ are in five different phases. Then the winding number can
be reliable topological quantum number to distinguish the quantum phases.

\textit{Conclusion. }In summary, a class of exactly solvable quantum Ising
models presented in this paper have obvious topological characterization and
indicate the existence of a topological quantum number, which is the winding
number for the loop in a two-dimensional auxiliary space and describes the
quantum phases in the extended quantum Ising model. This finding reveals the
connection between QPT and the geometrical order parameter characterizing
the phase diagram for a more generalized spin model, which will motivate
further investigation.

\acknowledgments We acknowledge the support of the National Basic Research
Program (973 Program) of China under Grant No. 2012CB921900 and the CNSF
(Grant No. 11374163).


\begin{thebibliography}{99}
\bibitem{Proof} State $\left\vert G_{2}^{\prime }\right\rangle $\ is the
ground state function of $h_{2}$ with eigenvalue $-1$ according to the exact
solution in Eq. (\ref{G_2_k}). The translational symmetry of $h_{2}$ ensures
the singlet $\left\vert G_{2}^{\prime }\right\rangle $\ to have $%
\left\langle G_{2}^{\prime }\right\vert \sigma _{j}^{z}\sigma
_{j-1}^{x}\sigma _{j+1}^{x}\left\vert G_{2}^{\prime }\right\rangle =-1$. On
the other hand, one can always rewrite state $\left\vert G_{2}\right\rangle $%
\ in the following form $\left\vert \mathrm{even}_{\mathrm{1}}\right\rangle
(\left\vert \uparrow \right\rangle _{j-1}\left\vert \uparrow \right\rangle
_{j}\left\vert \uparrow \right\rangle _{j+1}$\ $-\left\vert \downarrow
\right\rangle _{j-1}\left\vert \uparrow \right\rangle _{j}\left\vert
\downarrow \right\rangle _{j+1})$, $\left\vert \mathrm{even}_{\mathrm{2}%
}\right\rangle (\left\vert \downarrow \right\rangle _{j-1}\left\vert
\downarrow \right\rangle _{j}\left\vert \uparrow \right\rangle _{j+1}$ $%
+\left\vert \uparrow \right\rangle _{j-1}\left\vert \downarrow \right\rangle
_{j}\left\vert \downarrow \right\rangle _{j+1})$, $\left\vert \mathrm{odd}_{%
\mathrm{1}}\right\rangle (\left\vert \downarrow \right\rangle
_{j-1}\left\vert \uparrow \right\rangle _{j}\left\vert \uparrow
\right\rangle _{j+1}$ $-\left\vert \uparrow \right\rangle _{j-1}\left\vert
\uparrow \right\rangle _{j}\left\vert \downarrow \right\rangle _{j+1})$, $%
\left\vert \mathrm{odd}_{\mathrm{2}}\right\rangle (\left\vert \uparrow
\right\rangle _{j-1}\left\vert \downarrow \right\rangle _{j}\left\vert
\uparrow \right\rangle _{j+1}$ $+\left\vert \downarrow \right\rangle
_{j-1}\left\vert \downarrow \right\rangle _{j}\left\vert \downarrow
\right\rangle _{j+1})$, where $\left\vert \mathrm{even}\right\rangle $ and $%
\left\vert \mathrm{odd}\right\rangle $\ denote arbitrary states with even
and odd flips, respectively. And the indices $1$, $2$, denote the parity of
the summation which is the sum of the positions of the $\left\vert
\downarrow \right\rangle $. A straightforward algebra shows that $%
\left\langle G_{2}\right\vert \sigma _{j}^{z}\sigma _{j-1}^{x}\sigma
_{j+1}^{x}\left\vert G_{2}\right\rangle =-1$, which indicates $\left\vert
G_{2}^{\prime }\right\rangle =\left\vert G_{2}\right\rangle $, according to
the variational principle. Similarly, we have $\left\vert G_{-2}^{\prime
}\right\rangle =\left\vert G_{-2}\right\rangle $.
\end{thebibliography}
\end{document}